\documentclass[twocolumn]{emulateapj}

\usepackage{makecell}
\usepackage{amsmath}
\usepackage[utf8]{inputenc}
\usepackage{graphicx}
\usepackage{color}
\usepackage[colorlinks]{hyperref}
\hypersetup{
    colorlinks,	
    citecolor=blue,
}

\newcommand{\be}{\begin{eqnarray}}
\newcommand{\ee}{\end{eqnarray}}

\newcommand{\src}{GRS~1915+105}
\newcommand{\hxmt}{\textit{Insight}--HXMT}

\shorttitle{Disk wind in GRS~1915+105}
\shortauthors{Liu et al.}

\begin{document}

\title{The disk wind in GRS~1915+105 as seen by \textit{Insight}--HXMT}

\author{Honghui Liu\altaffilmark{1}, Yuhan Fu\altaffilmark{1}, Cosimo Bambi\altaffilmark{1, \dag}, Jiachen Jiang\altaffilmark{2}, Michael L. Parker\altaffilmark{2}, Long Ji\altaffilmark{3}, Lingda Kong\altaffilmark{4,5}, Liang Zhang\altaffilmark{4}, Shu Zhang\altaffilmark{4,5}, Yuexin Zhang\altaffilmark{6} }

\altaffiltext{1}{Center for Field Theory and Particle Physics and Department of Physics, 
Fudan University, 200438 Shanghai, China. \email[\dag E-mail: ]{bambi@fudan.edu.cn}} 
\altaffiltext{2}{Institute of Astronomy, University of Cambridge, Madingley Road, Cambridge CB3 0HA, UK} 
\altaffiltext{3}{School of Physics and Astronomy, Sun Yat-Sen University, 519082 Zhuhai, China}
\altaffiltext{4}{Key Laboratory for Particle Astrophysics, Institute of High Energy Physics, Chinese Academy of Sciences, 100049 Beijing, China}
\altaffiltext{5}{University of Chinese Academy of Sciences, Chinese Academy of Sciences, 100049 Beijing, China}
\altaffiltext{6}{Kapteyn Astronomical Institute, University of Groningen, P.O. BOX 800, 9700 AV Groningen, The Netherlands}

\begin{abstract}
We analyze three observations of GRS~1915+105 in 2017 by \textit{Insight}--HXMT when the source was in a spectrally soft state. We find strong absorption lines from highly ionized iron, which are due to absorption by disk wind outflowing at a velocity of $\sim$ 1000 km s$^{-1}$ along our line of sight. Two of the three observations show large amplitude oscillation in their light curves and the variation pattern corresponds to state $\kappa$ of GRS~1915+105. From time-averaged and flux-resolved analysis, we find that the variation of the ionization state of the disk wind follows the X-ray continuum on timescales from hundreds seconds to months. The radial location of the disk wind is consistent with thermal driving. The mass-loss rate due to the outflowing wind is comparable to the mass accretion rate in the inner disk, which demonstrates the important role of the disk wind in the disk accretion system.
\end{abstract}



\section{Introduction}


Ionized absorption winds have been commonly found in the X-ray spectra of low mass X-ray binary systems \citep[e.g.][]{Lee2002,  Ueda2010, Miller2015, Trigo2016, Neilsen2018}. These winds, in many observations, are found to be outflowing with respect to the central compact objects \citep[e.g.][]{Miller2006, Ueda2009, Blum2010, Neilsen2011, Neilsen2012, King2012, Miller2016}. Prominent features of the winds are blueshifted narrow absorption lines, e.g., \ion{Fe}{25} or \ion{Fe}{26}. The winds are more preferentially detected in high inclination systems \citep[see Fig.~2 of][]{Ponti2012}, which suggests an equatorial geometry and an origin from the accretion disk.

In the past two decades, we have learnt that the disk wind may be an important ingredient in the black hole accretion process. The amount of mass carried by the wind can be comparable or even significantly larger than the mass accretion rate at the inner disk \citep[e.g.][]{Ueda2009, Ponti2012}. The launch of disk winds might be related to state transitions of low mass X-ray binaries \citep[e.g.][]{Shields1986, Neilsen2011}. Observations that show anticorrelation between wind and jet \citep[e.g.][]{Neilsen2009} also indicate wind as an alternative mass ejection mode to quench the jet \citep[but see][]{Homan2016}. Moreover, there is obvious dependence of the detection of disk winds on the spectral states of black hole transients. Until now, most disk winds have been observed in the soft state \citep{Miller2008, Ponti2012} and the detections in the hard state have been very rare \citep[e.g.][]{Lee2002}. It remains to be known if this is because the wind is not launched (or not on the line of sight) in the hard state \citep[e.g.][]{Ueda2010, Neilsen2012b}. Another explanation could be that the wind is always launched, but it is too highly ionized \citep[e.g.][]{Shidatsu2019} or thermally unstable \citep[e.g.][]{Chakravorty2013, Petrucci2021} in the hard state.

Outflows cound be launched by thermal pressure \citep{Begelman1983, Done2018}, magnetic pressure \citep{Fukumura2017} or radiation pressure \citep[e.g.][]{Proga2002, Higginbottom2015}. However, it is not easy to discriminate different launching mechanisms due to the difficulty in determining the density and radial location of the disk wind. Moreover, different mechanisms may drive the disk winds at the same time and their relative importance may change with the accretion state \citep[e.g.][]{Neilsen2012b}.

Investigating the disk wind in low mass X-ray binaries (LMXRBs) may even help to understand the accretion environment of active galactic nuclei (AGNs). Recent studies have shown a possible connection between the broad line region (BLR) in AGNs and the stellar-mass black hole disk winds \citep[e.g.][]{Tremaine2014, Miller2015}. The wind may be relevant to the feedback to the environment if it carries a significant amount of momentum \citep[e.g.][]{King2013}. 

\src{} is a LMXRB that has played an important role in the study of the disk wind. It is one of the few sources that shows evidence of magnetically driven disk wind \citep[e.g.][]{Miller2016, Ratheesh2021}. A clear dichotomy of jets and disk winds is also present in this source \citep{Neilsen2009}. In some cases, the light curve of \src{} shows large amplitude variation on timescale from tens seconds to ks, which is ideal for the study of the connection between inner disk, outer disk and jet \citep[e.g.][]{Neilsen2011, Neilsen2012}. 

In this work, we analyze three observations (Fig.~\ref{log}) of \src{} by \hxmt{} \citep{Zhang2014} that show disk wind absorption signatures. The three observations show different variability patterns (see Fig.~\ref{lc}) but similar signatures of ionized absorption (see Fig.~\ref{resi}). With these data, we have the opportunity to study the disk wind, its variability on short (hundreds seconds) and long (months) timescales and its connection to different variability classes in \src{}.

\section{Observations and data reduction}\label{obs}

\begin{figure}
    \centering
    \includegraphics[width=\linewidth]{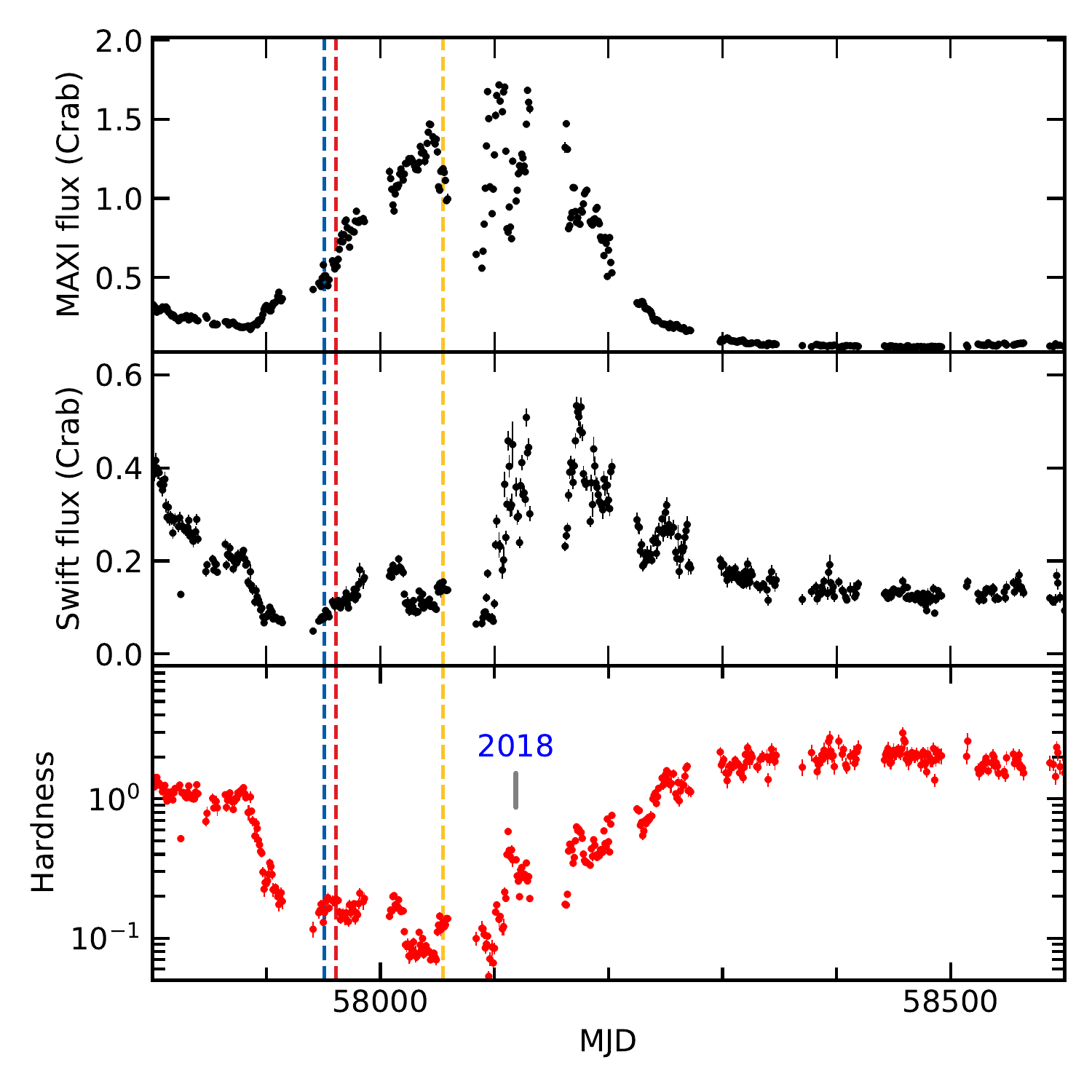}
    \caption{Light curves of \src{} in Crab units by MAXI/GSC (2--20 keV) and Swift/BAT (15--50 keV). The hardness in the lower panel is defined as the ratio between the Swift and MAXI count rate in Crab units (Swift/MAXI). The vertical lines mark the three \hxmt{} observations analyzed in this work.}
    \label{log}
\end{figure}

\hxmt{} observed \src{} three times when the source was in a ``soft state'' in 2017 (as shown in Fig.~\ref{log}). The source remained in a spectrally soft state for almost one year when there was a persistent disk wind \citep{Neilsen2018}. Details of the three \hxmt{} observations are shown in Tab.\ref{info-obs}.

\textit{Insight}-HXMT is the first Chinese X-ray telescope, which consists of low-energy (LE), medium-energy (ME) and high-energy (HE) detectors that cover the energy range of 1-250 keV \citep{Chen2020, Cao2020, Liu2020, Zhang2020}.  The lightcurves and spectra are extracted following the official user guide\footnote{\url{http://www.hxmt.cn/SoftDoc/67.jhtml}} and using the software {\sc HXMTDAS} ver 2.04. We estimate the background using standalone scripts \texttt{hebkgmap}, \texttt{mebkgmap} and \texttt{lebkgmap} \citep{Liao2020a, Guo2020, Liao2020b}. We screen good time intervals with the recommended criteria, i.e., the elevation angle $>$ 10 degree the geomagnetic cutoff rigidity $>$ 8 GeV, the pointing offset angle $<$ 0.1 and at least 300 s away from the South Atlantic Anomaly (SAA).

We fit data from \hxmt\ Low Energy X-ray Telescope (LE) in the energy range 2--9 keV and Medium Energy X-ray Telescope (ME) in 10--20 keV. Data from the High Energy X-ray Telescope (HE) are not included because of the low source count rate and high background. 

\begin{table*}
    \centering
    \renewcommand\arraystretch{2.0}
    \caption{\hxmt{} {\rm observations of \src{} analyzed in this paper}}
    \label{info-obs}
    \begin{tabular}{cccccc}
        \hline\hline
        Reference name & Date$^1$ & obsID & Exposure (ks) & Flux$^2$ & EW$^3$ (eV)\\
        \hline
        H1 & 20170717(18) & P0101330001 & 26.6 & $1.192\pm0.01$  & $20.8_{-1.9}^{+1.4}$ \\ 
        \hline
        H2 & 20170727 & P0101310001 & 10.4 & $1.513\pm0.03$  & $16.6_{-2.7}^{+2.0}$ \\
        \hline
        H3 & 20171029(30) & P0101310002 & 23.3 & $2.583\pm0.01$  & $27.7_{-1.3}^{+1.2}$\\
        \hline

    \end{tabular}\\

\textit{Note}. (1) The observation date is presented in the form of \texttt{yyyymmdd}. (2) The observed flux (in units of 10$^{-8}$ erg cm$^{-2}$ s$^{-1}$ in 2--10 keV) is calculated from the best-fit model in XSPEC. (3) The equivalent width of the \ion{Fe}{26} Ly$\alpha$ absorption line.
\end{table*}


\section{Data analysis}\label{s-ana}

The spectra of \src{} are analyzed with XSPEC v12.10.1f~\citep{xspec}. The cross section is set to~\cite{Verner1996} and the element abundances to~\cite{Wilms2000}. In this manuscript, the uncertainties are quoted at 90\% confidence level. 

\subsection{The light curve}

\begin{figure*}
    \centering
    \includegraphics[width=\linewidth]{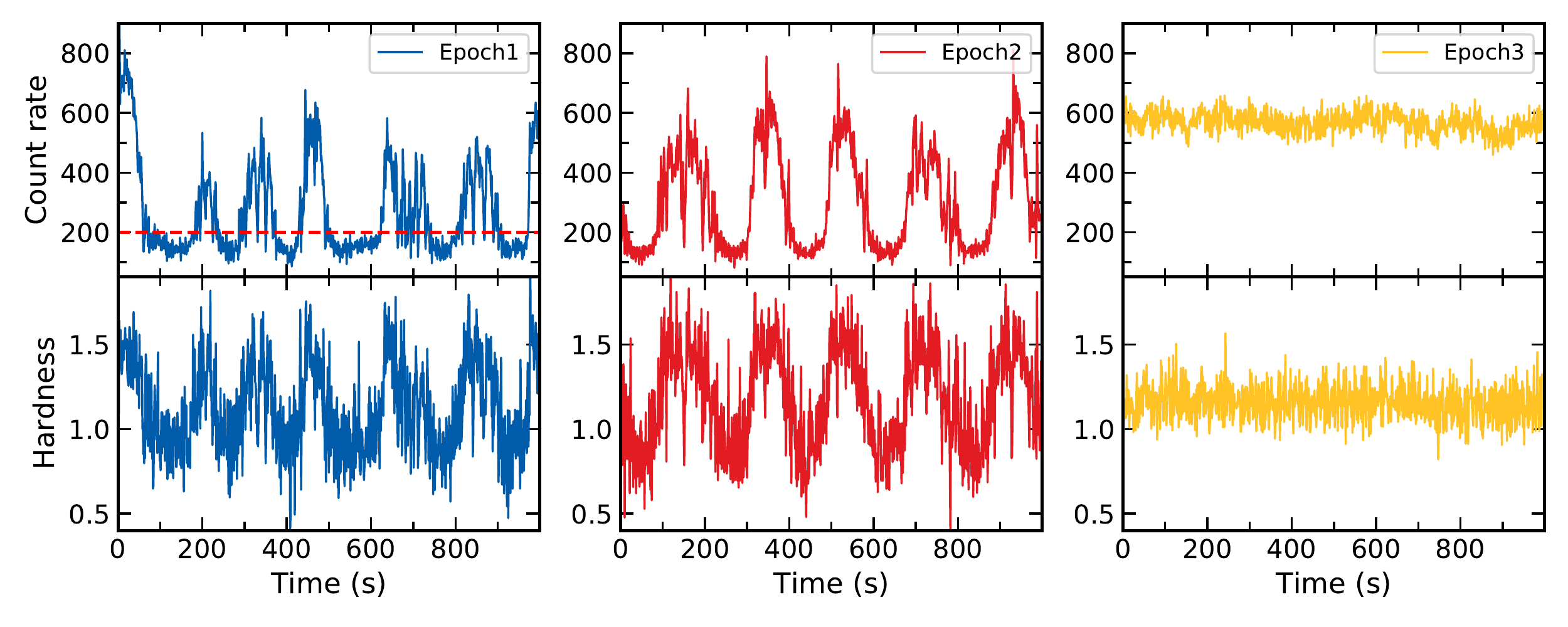}
    \caption{The LE (1--10 keV) light curve and the hardness ratio of \src{} for the three epochs. For every epoch, only 1000~s (bin size 1~s) data are shown. The hardness ratio is defined as ratio between the count rates in the range of 4--10 keV and 1--4 keV. The red dashed line in the panel for Epoch 1 denotes the threshold used to divide this observation to low and high flux states (see Sec.~\ref{Short-time}).}
    \label{lc}
\end{figure*}

GRS 1915+105 is a particular source since it does not go frequently into outbursts or trace certain patterns on the hardness intensity diagram (HID) as other canonical  LMXRBs (e.g. GX 339-4). The source has remained bright since its discovery in 1992 by WATCH~\citep{Castro-Tirado1992}. The light curve of \src{} can be divided into 14 classes, according to its variability pattern and color-color diagram (CCD)~\citep{Belloni2000, Klein-Wolt2002, Hannikainen2005}. We show in Fig.~\ref{lc} typical light curves of the three observations. For epoch 1 and 2, the LE count rate oscillates with a large amplitude (more than a factor of 3) on a time scale of 200 seconds. Epoch 3, however, does not show large amplitude variability. Comparing the light curves and CCD (not shown here) to \cite{Belloni2000}, we find that epoch 1 and 2 belong to class $\kappa$ and epoch 3 to class $\delta$. 

\subsection{The continuum and absorption lines}
\label{continuum}

\begin{figure*}
    \centering
    \includegraphics[width=0.49\linewidth]{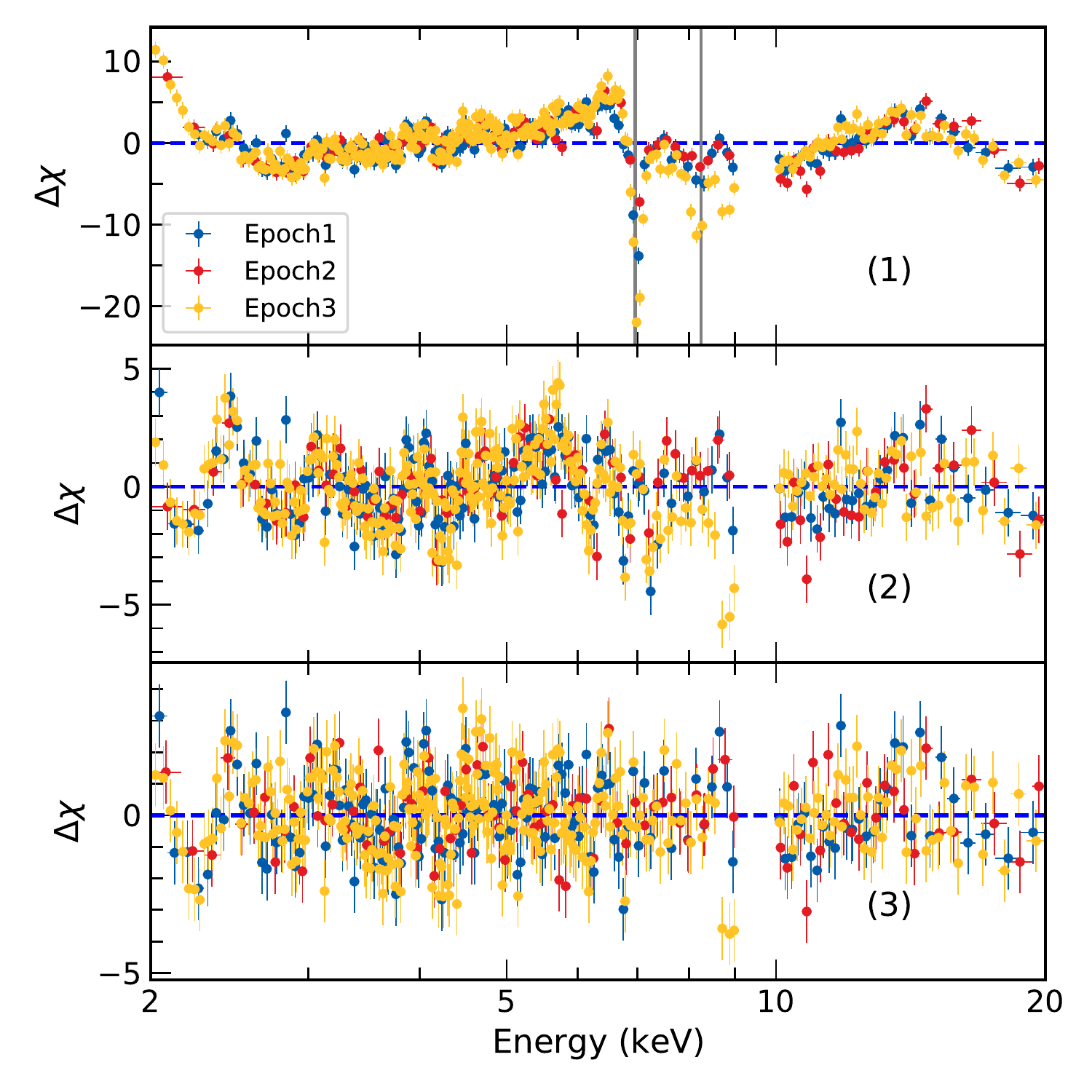}
    \includegraphics[width=0.49\linewidth]{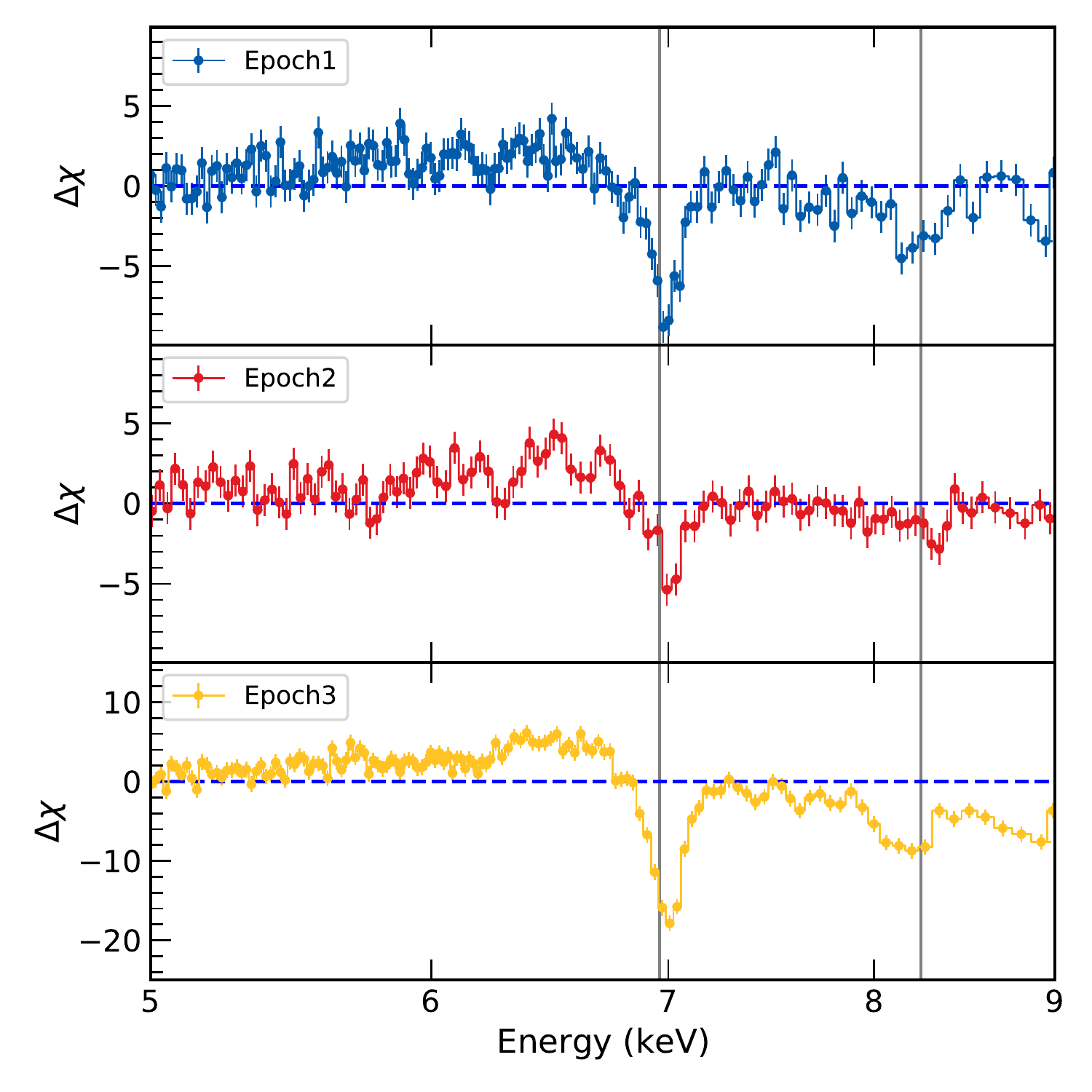}
    \caption{(Left 1): Residuals to an absorbed continuum model \texttt{tbabs*(diskbb + nthcomp)} for the three epochs. The two vertical grey lines mark the positions of absorption by \ion{Fe}{26} Ly$\alpha$ and \ion{Fe}{26} Ly$\beta$. (Left 2): Residuals to the model \texttt{tbabs*(diskbb + nthcomp + relxill)} with absorption lines fitted with two extra negative \texttt{gaussian}. (Left 3): Residuals to the model \texttt{tbabs*pcfabs*(nthcomp + gaussian)} with two additional absorption gaussian lines. (Right): A zoomed in version of panel 1 on the left with the three observations plotted separately. Data are rebinned for visual clarity.}
    \label{resi}
\end{figure*}

\textit{Model 1.} As the first step, we fit the broadband spectra with a multicolor disk component (\texttt{diskbb}, \cite{Mitsuda1984}) plus a power-law component \citep[\texttt{nthcomp},][]{Zdziarski1996, Zycki1999}. The model \texttt{tbabs} is also included to model the absorption by interstellar medium (ISM), with its column density tied across all observations. The residuals to the best-fit are shown in the top panel of Fig.~\ref{resi}, from which we can identify the missing components in the current model. We can see an excess below 3 keV and a hump in the iron band. There is a deep trough at 7 keV, suggesting the presence of absorption by highly ionized iron (e.g. \ion{Fe}{26} Ly$\alpha$). Absorption by \ion{Fe}{26} Ly$\beta$ (8.25 keV) is also present with a significance of 4$\sigma$ for epoch 1 and 3. Moreover, the \ion{Fe}{26} 1s--4p absorption line (8.7~keV) is only significant in epoch 3 (3$\sigma$).

\textit{Model 2.} The hump peaked around 6.4 keV in Fig.~\ref{resi} indicates the existence of a reflection component by the optically thick accretion disk surrounding the black hole, which is commonly seen in the X-ray spectra of black hole X-ray binaries~\citep[e.g.][]{Fabian1989, Miller2013, Walton2016, Jiang2019, Jiang2020}. So we add a relativistic reflection component (\texttt{relxill}, \cite{Garcia2014}) to the model with two negative narrow ($\sigma$=10 eV) gaussian lines to fit the absorption by highly ionized iron. This model gives an acceptable fit ($\chi^2/\nu$=3564/2906), although there is still some curvature left in the residuals at 5--6 keV (see the middle panel of Fig.~\ref{resi}).


\textit{Model 3.} In the end, we note that partially covered absorbers have been applied to fit the X-ray spectra of \src{} in its flaring events \citep[e.g.][]{Neilsen2020, Kong2021}. We also include a partial covering absorption component (\texttt{pcfabs}) to fit the three spectra. In this fit, we find that the \texttt{diskbb} component is not needed to fit the continuum since the fit improves only by $\Delta\chi^2 \sim 3$ with respect to the model without this component. In addition, a gaussian line centered at 6.4~keV instead of a \texttt{relxill} component is already enough to fit the reflection features in the spectra. In XSPEC notation, the model used here is \texttt{tbabs*pcfabs*(nthcomp+gauss+gauss2+gauss3)}, where \texttt{gauss2} and \texttt{gauss3} are two negative gaussians to fit the iron absorption lines. This model provides a better fit statistics ($\chi^2/\nu$=3138/2911) than the relativistic reflection model and shows less features in the residuals at 5--6 keV (see Fig.~\ref{resi}). The equivalent width (EW) of the \ion{Fe}{26} Ly$\alpha$ absorption line measured with this model is shown in Tab.~\ref{info-obs} and it indicates that the strongest absorption is on epoch 3.

Note that the broadband X-ray spectra of the three observations are complex and it is not the aim of this paper to study their nature. We fit the broadband X-ray spectra to reveal the absorption features by disk wind and determine the model that can best describe the continuum. The best-fit parameters of the continuum model should be treated as phenomenological results.

\begin{figure}
    \centering
    \includegraphics[width=\linewidth]{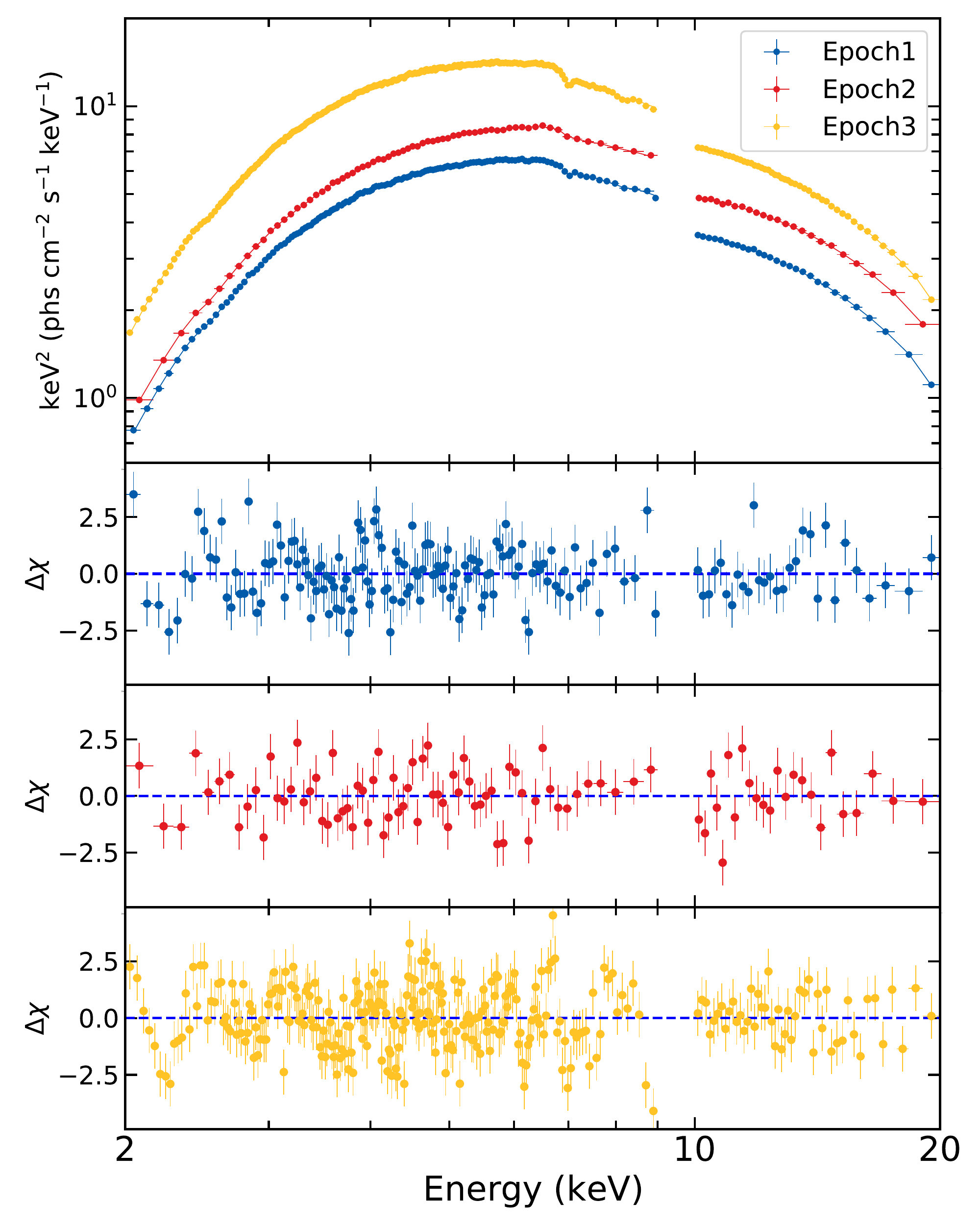}
    \caption{Spectra and residuals for the best-fit model \texttt{tbabs*xstar*pcfabs(nthcomp + gauss)} for the three epochs. The spectra are corrected for the effective area of the instruments but not unfolded from the response. Data are rebinned for visual clarity.}
    \label{eeuf}
\end{figure}

\subsection{The absorption wind}
\label{xstar}
\subsubsection{Time-averaged analysis}

To obtain the physical properties (e.g., ionization state, column density) of the disk wind from the absorption signatures, we calculate a grid model with XSTAR (version 2.2) code~\citep{Kallman2001, Kallman2004} assuming the absorption corrected best-fit continuum model from \textit{Model 1} (see Sec.~\ref{continuum}). The turbulent velocity is set to be 1000~km~s$^{-1}$ and density ($n$) to 10$^{12}$~cm$^{-3}$. The grid model is applied to the three observations with the continuum being modelled with partially absorbed corona emission. The final model in XSPEC notation is: \texttt{tbabs*xstar*pcfabs*(nthcomp+gauss)} (\textit{Model 4}). We fit the three observations simultaneously with the column density ($N_{\rm H}$) of \texttt{tbabs} and the cross normalization constant between LE and ME tied across observations. This single zone absorption model already gives an acceptable fit ($\chi^2$/d.o.f=3201/2914). 

We also test if a second layer of absorption is needed to fit the spectra. Adding a second XSTAR layer improves the $\chi^2$ by 70 with nine more degrees of freedom. However, we find that the column density for epoch 1 and 2 of the second layer is pegged at the lower limit and has negligible effect on the continuum. The outflowing velocity and ionization state are also not constrained. The improvement on $\chi^2$ is mainly due to the spectrum of epoch 3. Therefore, we add this second layer only for epoch 3 and keep only one layer of absorption for epoch 1 and 2. We report the best-fit parameters for both the single zone and two zones model in Tab.~\ref{bestfit}, so it is straightforward to assess the change of parameters by including a second absorption zone. The results in Tab.~\ref{bestfit} show that the second absorber of epoch 3 has a very high outflow velocity ($\sim$ 0.05 c). However, a detailed investigation of the absorption lines produced by the second layer reveals that it is still fitting the absorption line at 7~keV. The high outflow velocity simply shifts the energy of \ion{Fe}{25} (6.70 keV) line to 7~keV. Multiple absorption zones have also been found in other LMXRBs~\citep[e.g.][]{Miller2015}. The higher velocity components are usually found to be more highly ionized, which is the opposite to our fit. The results of the two zones model should thus be taken with caution and we show the unfolded spectra and residuals for only the single zone model in Fig.~\ref{eeuf}. We will also focus our discussion on the single zone model.

From the best-fit parameters of the single-zone model, we can see that the wind is outflowing from the X-ray source with velocity $\sim$ 1000~km~s$^{-1}$. There are also apparent variations of column density of the disk wind. Epoch 3 has the largest column density as already suggested by the analysis of the equivalent width of the absorption line. In addition, the absorption corrected ionizing flux increases by a factor of 2 from epoch 1 to epoch 3. If we assume a constant density and location of the disk wind, the definition of ionization parameter ($\xi=L/nr^2$) would indicate an increase of the ionization parameter by $\Delta \log(\xi)=0.3$. The variation of the ionization parameter we measured is in agreement with this value, although the upper error bar for epoch 1 is relatively large (see Tab.~\ref{bestfit}). This is even true for the two zone model but the large uncertainty on the ionization parameter for epoch 3 prevents any compelling conclusion.

\begin{table*}
    \centering
    \caption{Best-fit values.}
    \label{bestfit}
    \renewcommand\arraystretch{2.0}
    \begin{tabular}{lc|ccc|ccc}
        \hline\hline
                        & & \multicolumn{3}{c|}{Sigle zone} & \multicolumn{3}{c}{Two zones} \\
        Component       & Parameter & Epoch 1 & Epoch 2 & Epoch 3 & Epoch 1 & Epoch 2 & Epoch 3 \\
        \hline
        \textsc{tbabs}  & $N_{\rm H}$ (10$^{22}$ cm$^{-2})$ &  & $6.21_{-0.24}^{+0.09}$ &  &  & $6.01_{-0.21}^{+0.17}$ & \\
        \hline
        \textsc{xstar}  & $N_{\rm H}$ (10$^{22}$ cm$^{-2})$ & $7.9_{-1.1}^{+1.6}$ & $6.5_{-1.3}^{+2.1}$ & $12.7_{-0.3}^{+0.4}$ & $7.9_{-1.0}^{+1.5}$ & $6.4_{-1.3}^{+2.1}$ & $10.6_{-4.0}^{+2.1}$ \\
                        & $\log(\xi)$ & $4.28_{-0.07}^{+0.2}$ & $4.33_{-0.13}^{+0.26}$ & $4.623_{-0.029}^{+0.05}$ & $4.28_{-0.06}^{+0.18}$ & $4.33_{-0.12}^{+0.24}$ & $4.69_{-0.4}^{+0.15}$ \\
                        & $v$ (km s$^{-1}$) & $900_{-300}^{+400}$ & $1200_{-700}^{+700}$ & $1300_{-200}^{+200}$ & $900_{-300}^{+400}$ & $1300_{-700}^{+700}$ & $700_{-500}^{+500}$ \\
        \hline
        \textsc{xstar}  & $N_{\rm H}$ (10$^{22}$ cm$^{-2})$ & - & - & - & - & - & $0.77_{-0.24}^{+0.20}$ \\
                        & $\log(\xi)$ & - & - & - & - & - & $3.40_{-0.12}^{+0.24}$ \\
                        & $v$ (km s$^{-1}$) & - & - & - & - & - & $15000_{-1000}^{+1200}$ \\
        \hline
        \textsc{pcfabs} & $N_{\rm H}$ (10$^{22}$ cm$^{-2})$ & $11.6_{-0.9}^{+0.6}$ & $11.9_{-0.7}^{+0.6}$ & $11.9_{-1.0}^{+0.5}$ & $10.6_{-0.8}^{+0.8}$ & $11.1_{-0.7}^{+0.7}$ & $10.6_{-0.8}^{+0.8}$ \\
                        & $f_{\rm c}$ & $0.374_{-0.017}^{+0.04}$ & $0.459_{-0.018}^{+0.04}$ & $0.369_{-0.017}^{+0.03}$ & $0.41_{-0.03}^{+0.04}$ & $0.472_{-0.022}^{+0.04}$ & $0.393_{-0.027}^{+0.04}$ \\
        \hline
        \textsc{nthcomp} & $\Gamma$ & $2.401_{-0.05}^{+0.026}$ & $2.293_{-0.023}^{+0.009}$ & $2.632_{-0.026}^{+0.017}$ & $2.40_{-0.03}^{+0.05}$ & $2.302_{-0.026}^{+0.06}$ & $2.64_{-0.03}^{+0.04}$ \\
                        & $kT_{\rm e}$ (keV) & $3.34_{-0.04}^{+0.04}$ & $3.34_{-0.04}^{+0.04}$ & $3.69_{-0.04}^{+0.05}$ & $3.34_{-0.06}^{+0.04}$ & $3.36_{-0.04}^{+0.04}$ & $3.73_{-0.05}^{+0.1}$ \\
                        & $kT_{\rm bb}$ (keV) & $1.07_{-0.07}^{+0.09}$ & $0.94_{-0.06}^{+0.18}$ & $1.12_{-0.03}^{+0.06}$ & $1.07_{-0.07}^{+0.06}$ & $0.98_{-0.1}^{+0.13}$ & $1.135_{-0.019}^{+0.04}$ \\
                        & Norm & $3.26_{-0.12}^{+0.13}$ & $4.6_{-0.4}^{+0.4}$ & $7.52_{-0.2}^{+0.14}$ & $3.23_{-0.22}^{+0.25}$ & $4.32_{-0.25}^{+0.6}$ & $7.32_{-0.25}^{+0.19}$ \\
        \hline
        \textsc{gauss}  & $E_{\rm line}$ (keV) & $6.4^*$ & $6.4^*$ & $6.4^*$ & $6.4^*$ & $6.4^*$ & $6.4^*$ \\
                        & EW (keV) & $0.17_{-0.05}^{+0.05}$ & $0.29_{-0.05}^{+0.07}$ & $0.101_{-0.016}^{+0.03}$ & $0.17_{-0.05}^{+0.05}$ & $0.29_{-0.06}^{+0.06}$ & $0.101_{-0.016}^{+0.025}$\\
                        & Norm (10$^{-2}$ Phs cm$^{-2}$ s$^{-1}$) & $3.0_{-0.4}^{+0.3}$ & $6.9_{-0.6}^{+0.5}$ & $4.1_{-0.3}^{+0.4}$& $3.1_{-0.6}^{+0.3}$ & $6.6_{-1.2}^{+0.7}$ & $4.1_{-0.4}^{+0.5}$ \\
        \hline
                        & $\chi^2$/d.o.f & & 3201.4/2914 & & & 3153.7/2911 & \\
        \hline\hline
    \end{tabular}

    \textit{Note.} Best-fit parameters for the model \texttt{tbabs*xstar*pcfabs(nthcomp + gauss)}. The \texttt{gauss} component is used to fit possible reflected emission from the accretion disk. Note that there is a second layer of \texttt{xstar} for epoch 3 in the ``Two zones" case. Parameters with $^*$ are fixed during the fit.
    \vspace{0.5cm}
\end{table*}

\subsubsection{Flux-resolved analysis}
\label{f-resolve}

The light curve in Fig.~\ref{lc} shows that the count rate of \src{} varies by a factor of 3 on a timescale of $\sim 200$ s for epoch 1 and 2. It is of importance to see if the physical properties (e.g., the ionization parameter) of the disk wind respond to this fast variation of the ionizing continuum. Since epoch 1 has the longest exposure, we divide this observation into low flux and high flux states with a threshold of 200 counts/s for LE (the red line in Fig.~\ref{lc}). 

We first fit the flux-resolved spectra with \textit{Model 2} to obtain the equivalent width of the \ion{Fe}{26} Ly$\alpha$ absorption line in the low and high flux intervals. The result shows that the EW evolves from $21.4_{-2.6}^{+3.0}$ eV for the low flux state to $15.9_{-0.9}^{+2.6}$ eV for the high flux state. This suggests a lower ionic column density of \ion{Fe}{26} Ly$\alpha$ in the high flux state, which could be due to change of the line of sight column density of the wind or to the fact that more iron ions are fully stripped due to stronger ionizing flux in the high state. The degeneracy between the two scenarios may lead to parameter degeneracy when fitting with a photoionization model.

\begin{figure}
    \centering
    \includegraphics[width=\linewidth]{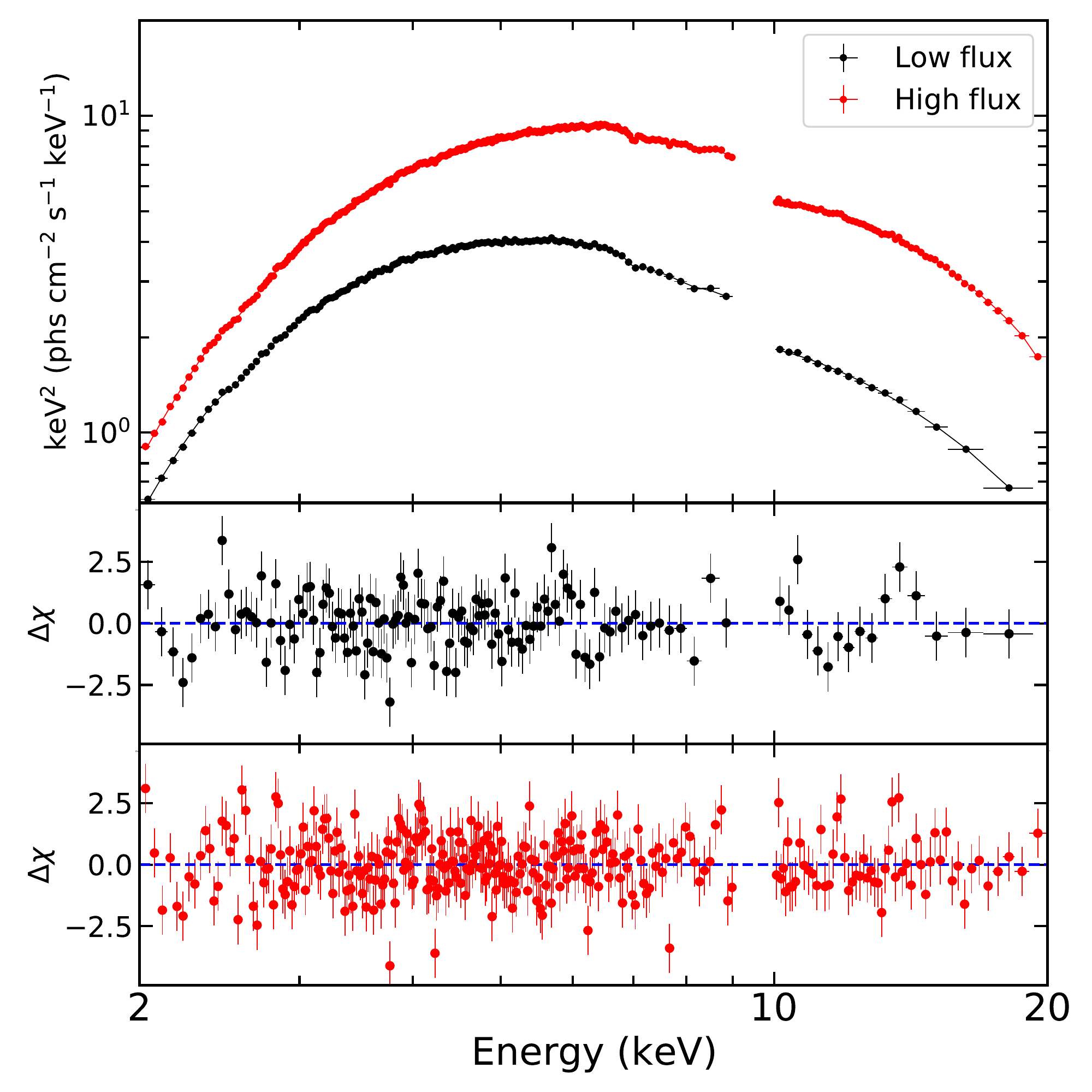}
    \caption{Spectra and residuals for the low and high flux data of epoch 1. The spectra are corrected for the effective area and are only shown for demonstration purposes. Data are rebinned for visual clarity.}
    \label{flux_eeuf}
\end{figure}

We then fit both spectra simultaneously with \textit{Model 4} using new XSTAR grids calculated with the low and high flux spectral continuum. We note that, even if there is strong variation of the ionizing flux, it is still a safe assumption that the absorption gas is in ionization equilibrium since the recombination timescale can be well below 1~s for a typical density of the wind~\citep{Kallman2009}. In this case, if we leave the column density and outflowing velocity of XSTAR free to vary between the two spectra, we obtain consistent results ($N_{\rm H}$ is $8.4_{-2.5}^{+2.4}\times 10^{22}$ cm$^{-2}$ in low flux state and $6.4_{-1.5}^{+2.7}\times 10^{22}$ cm$^{-2}$ in the high flux state). The ionization parameter ($\log(\xi)$) is only loosely constrained (low: $4.18_{-0.15}^{+0.09}$, high: $4.05_{-0.12}^{+0.14}$) possibly due to the degeneracy discussed above. We also consider the possibility that we are seeing the same wind in low and high flux states by linking the column density and velocity of the XSTAR model. This is a reasonable assumption since the traveling timescale of the wind is much longer than the timescale of the cycle ($\sim$ 200~s) we are considering \citep[see][]{Ueda2010}. The change of fit statistics is also minor ($\Delta\chi^2=1$) compared to the last case when $N_{\rm H}$ and $v$ are not linked between the two spectra. The results of this model are shown in Tab.~\ref{low-high} and Fig.~\ref{flux_eeuf}.

\begin{table}
    \centering
    \caption{Best-fit values.}
    \label{low-high}
    \renewcommand\arraystretch{2.0}
    \begin{tabular}{lccc}
        \hline\hline
        Component       & Parameter & Low Flux & High Flux  \\
        \hline
        \textsc{tbabs}  & $N_{\rm H}$ (10$^{22}$ cm$^{-2})$ & \multicolumn{2}{c}{$5.6_{-0.8}^{+0.4}$}   \\
        \hline
        \textsc{xstar}  & $N_{\rm H}$ (10$^{22}$ cm$^{-2})$ &  \multicolumn{2}{c}{$7.6_{-2.0}^{+1.5}$}  \\
                        & $\log(\xi)$ & $4.15_{-0.14}^{+0.08}$ & $4.12_{-0.13}^{+0.08}$  \\
                        & $v$ (km s$^{-1}$) & \multicolumn{2}{c}{$1000_{-360}^{+360}$}  \\
        \hline
        \textsc{pcfabs} & $N_{\rm H}$ (10$^{22}$ cm$^{-2})$ & $9.7_{-2.4}^{+2.5}$ & $7.9_{-1.3}^{+1.4}$  \\
                        & $f_{\rm c}$ & $0.39_{-0.08}^{+0.14}$ & $0.45_{-0.09}^{+0.15}$  \\
        \hline
        \textsc{nthcomp} & $\Gamma$ & $2.76_{-0.10}^{+0.11}$ & $2.35_{-0.13}^{+0.16}$  \\
                        & $kT_{\rm e}$ (keV) & $3.76_{-0.22}^{+0.36}$ & $3.33_{-0.15}^{+0.23}$  \\
                        & $kT_{\rm bb}$ (keV) & $1.08_{-0.09}^{+0.09}$ & $1.35_{-0.23}^{+0.15}$ \\
                        & Norm & $2.38_{-0.20}^{+0.24}$ & $3.23_{-0.23}^{+0.57}$  \\
        \hline
        \textsc{gauss}  & $E_{\rm line}$ (keV) & $6.4^*$ & $6.4^*$  \\
                        & EW (keV) & $0.046_{-0.029}^{+0.029}$ & $0.08_{-0.05}^{+0.05}$  \\
                        & Norm  & $0.50_{-0.18}^{+0.26}$ & $1.8_{-1.3}^{+2.1}$ \\
                        & (10$^{-2}$ Phs/cm$^{2}$/s) & & \\
        \hline
                        & $\chi^2$/d.o.f &  \multicolumn{2}{c}{2024/1904} \\
        \hline\hline
    \end{tabular}

    \textit{Note.} Best-fit parameters for the low and high flux state spectra with \textit{Model 4}. Parameters with $^*$ are fixed during the fit.
\end{table}

\section{Discussion and conclusions}\label{s-dis}

\subsection{Long timescale variability of the disk wind}

In the three soft spectra of \src{}, we find the presence of highly ionized ($\log(\xi)>$4.2) and high column ($N_{\rm H}\sim$ 10$^{23}$ cm$^{-2}$) absorption wind along the line of sight. On a timescale of three months, we find that the ionization state of the disk wind is following the changes of the ionizing flux, suggesting a photoionization origin. We note that \cite{Neilsen2018} find a persistent accretion disk wind in \src{} using \textit{NICER} observations from 2017 June to November, which cover well the three observations we are studying. \cite{Neilsen2018} find the \ion{Fe}{25} absorption line is less common than \ion{Fe}{26} and the ionization parameter $\xi$ should be larger than 10$^4$ from simple line ratio analysis. These results are confirmed by our analysis of \hxmt{} data. The other interesting property found by \cite{Neilsen2018} is that the column density of \ion{Fe}{26} is fairly steady across observations in the five months, which is attributed to either a steady of total column density or the balance between variations of total column density and ionization state. We find that, from epoch 1 to epoch 3, there is clear change of the total column density and ionization state (single zone model) of the absorber to support the latter scenario. The tentative evidence of overionization in the study of short timescale variability of the wind (see Sec.~\ref{Short-time}) gives more support on this scenario. It is possible that, as the X-ray irradiation on the outer disk becomes stronger, more material is puffed up from the disk to the line of sight. However, the stronger ionizing flux also gets the wind more ionized, which in the end maintains a relatively steady ionic column of \ion{Fe}{26}.

Epoch 3 exhibits a different variability pattern from that of epoch 1 and 2. However, we do not find any consequence of this difference on the dynamics of the disk wind. The higher column density and ionization parameter in epoch 3 seems merely a result of stronger X-ray illumination. When considering a two zone model, we find a fast absorber in epoch 3 outflowing at $\sim$ 0.05~c. However, as we noted, the detection of this fast absorber is not convincing enough, since it is mainly fitting the same line (7~keV) as the slow component and there are no other detectable signatures.

\subsection{Short timescale variability of the disk wind}
\label{Short-time}

The flux-resolved spectra allow us to study the variability of the disk wind on a timescale of hundreds of second. The spectra in Fig.~\ref{flux_eeuf} show that the source is harder in the higher flux state. Spectral fitting also reveals a harder powerlaw emission in the high state (Tab.~\ref{low-high}). 

A preliminary line width analysis reveals that the change in equivalent width of the \ion{Fe}{26} line is marginally significant at 90\% confidence (see Sec.~\ref{f-resolve}). The drop in equivalent width in the high flux suggests that the wind gets more ionized as the X-ray luminosity increases. As we noted, the source spectrum is harder in the higher flux state, which means there are more high energy photons to fully strip the \ion{Fe}{26} ions.

Variability of the disk wind in \src{} on short timescales (from seconds to ks) has been studied with a flux-resolved strategy in $\rho$, $\beta$, $\gamma$ and $\theta$ classes~\citep[e.g.][]{Ueda2010, Neilsen2011, Neilsen2012}. To our best knowledge, this work is the first time that the variability of the disk wind is studied in $\kappa$ state. Our finding is consistent with \cite{Ueda2010}, where the absorption line is weaker when the hard X-ray flux is stronger. This trend clearly suggests photoionization of the disk wind.  In \cite{Lee2002}, the variation amplitude of the ionization parameter of the disk wind is much higher than that of the ionizing flux, for which the authors claimed there was change of density of the disk wind on timescales of ks. Based on the same argument, \cite{Neilsen2011} find density variation on timescale of 5~s in the $\rho$ state. From the flux-resolved analysis, we do not find any evidence of variation of the ionization parameter when the source flux changes (Tab.~\ref{low-high}). This suggests that we are seeing change of the density of the disk wind on a timescale of 200~s.

\subsection{Driving mechanism and mass outflow}
\label{mechanism}

Accretion disk wind can be driven by radiation, thermal or magnetic pressure. The high ionization state we find already excludes the radiation pressure as the dominant mechanism since radiation pressure depends mostly on UV absorption lines and is only effective when $\log(\xi)<3$ \citep{Proga2002, Proga2003}. The thermally driven wind can only be launched at relatively large distances ($\sim 10^{4}~r_{\rm g}$, \cite{Begelman1983, Woods1996}). It is thus of crucial importance to locate the wind in order to understand its driving mechanism.

The upper limit of the wind location can be estimated by geometric consideration. Since the thickness of the wind $\Delta R$ can not exceed its distance to the central black $r$, we would have:
\begin{equation}
r<\frac{L_{\rm ion}}{N_{\rm H}\xi}
\end{equation}
if assuming $N_{\rm H}=n\Delta R$ and including the definition of ionization parameter. In this way, the maximum of the radial location will be $8\times10^4 R_{\rm g}$, $11\times10^4 R_{\rm g}$ and $5\times10^4 R_{\rm g}$ for absorbers of epoch 1, 2 and 3, respectively. A lower limit can be placed if we assume that the outflow velocity of the wind is larger than the local escape velocity: 
\begin{equation}
r_{\rm min} = 2GM_{\rm BH}v^{-2}_{\rm out}
\end{equation}
This would give a lower boundary of $11\times10^4 R_{\rm g}$, $5\times10^4 R_{\rm g}$ and $8\times10^4 R_{\rm g}$ for epoch 1, 2 and 3 for a black hole mass of 12.4 $M_{\odot}$ \citep{Reid2014}. We note that this is only a loose constraint. The wind can be much closer than this lower boundary if a substantial amount of its velocity is perpendicular to the line of sight. Therefore, the lower limit can be further relaxed to be reconciled with the upper limit. These estimates indicate a launching radius about $10^4 \sim 10^{5} R_{\rm g}$ for the disk wind. This range is consistent with that predicted by a thermally driven wind and magnetic pressure is not needed.

We can also estimate the mass outflow rate of the disk wind via:
\begin{equation}
    \dot{M}_{\rm wind}=f\mu4\pi r^2nm_{\rm p}v\frac{\Omega}{4\pi} = f\mu\Omega m_{\rm p}v(\frac{L_{\rm ion}}{\xi})
\end{equation}
where $\Omega/4\pi$ is the covering factor, $f$ is volume filling factor, $m_{\rm p}$ is the proton mass and $\mu$ is the mean atomic weight ($\mu=1.23$ with solar abundances). Assuming the parameters of epoch 3, this gives 
\begin{equation}
    \dot{M}_{\rm wind}=3.1\times 10^{19} f (\frac{\Omega}{4\pi}) ~\rm{g}~\rm{s}^{-1}
\end{equation}
The mass accretion rate can be estimated as $M_{\rm acc} = L/\eta c^{2} = 5.6\times 10^{18}$ g s$^{-1}$, where the efficiency $\eta=0.1$. This comparison shows that, as the covering factor and volume filling factor approach unity, the mass outflow carried by the wind can be comparable or even larger than the mass accretion rate in the inner disk. The importance of disk wind in the understanding of accretion process is obvious.


{\bf Acknowledgments --} The work of H.L. and C.B. is supported by the Natural Science Foundation of Shanghai, Grant No. 22ZR1403400, the Shanghai Municipal Education Commission, Grant No. 2019-01-07-00-07-E00035, the National Natural Science Foundation of China (NSFC), Grant No. 11973019, and Fudan University, Grant No. JIH1512604. J.J. acknowledges support from the Leverhulme Trust, the Isaac Newton Trust and St Edmund's College, University of Cambridge. Y.Z. acknowledges support from China Scholarship Council (CSC 201906100030). S.Z. thanks the support from the National Key R\&D Program of China (2021YFA0718500) and the National Natural Science Foundation of China, Grant No. U1838201, U1838202.



\bibliographystyle{apj}
\bibliography{bibliography}


\end{document}